\input{aipcheck}

\documentclass[
    ,final            % use final for the camera ready runs
  ]
  {aipproc}

\layoutstyle{6x9}
\usepackage{axodraw}
\begin{document}

\title{Neutrino Mass Seesaw Version 3:\\ Recent Developments}


\author{Ernest Ma}{
  address={Physics and Astronomy Department, University of California, 
Riverside, California 92521, USA}
}

\begin{abstract}
The origin of neutrino mass is usually attributed to a seesaw mechanism, 
either through a heavy Majorana fermion singlet (version 1) or a heavy 
scalar triplet (version 2).  Recently, the idea of using a heavy Majorana 
fermion triplet (version 3) has gained some attention.  This is a review 
of the basic idea involved, its U(1) gauge extension, and some recent 
developments.
\end{abstract}

\maketitle

%%%%%%%%%%%%%%%%%%%%%%%%%%%%%%%%%%%%%%%%%%%%
%% MAINMATTER
%%%%%%%%%%%%%%%%%%%%%%%%%%%%%%%%%%%%%%%%%%%%

\section{Introduction}

In the minimal standard model (SM) of quarks and leptons, the neutrinos 
$\nu_{e,\mu,\tau}$ are very different from other fermions because they need 
only exist as the neutral components of the electroweak doublets $L_\alpha = 
(\nu_\alpha,l_\alpha)$.  As such, they are massless two-component spinors 
and may become massive only if there is new physics beyond the SM.  
Assuming only the low-energy particle content of the SM, it was pointed 
out long ago \cite{w79} that small Majorana neutrino masses are given by 
the unique dimension-five operator
\begin{equation}
{\cal L}_5 = {f_{\alpha \beta} \over 2 \Lambda} (\nu_\alpha \phi^0 - l_\alpha 
\phi^+) (\nu_\beta \phi^0 - l_\beta \phi^+),
\end{equation}
where $\Phi = (\phi^+,\phi^0)$ is the one Higgs scalar doublet of the SM. 
The neutrino mass matrix is thus necessarily seesaw in form, i.e. 
$f_{\alpha \beta} v^2/\Lambda$, where $v$ is the vacuum expectation value 
of $\phi^0$ which breaks the electroweak $SU(2) \times U(1)$ gauge symmetry.  
It was also pointed out some years ago \cite{m98} that there are three (and 
only three) tree-level realizations of this operator (Fig.~1), as well as 
three generic one-loop realizations.
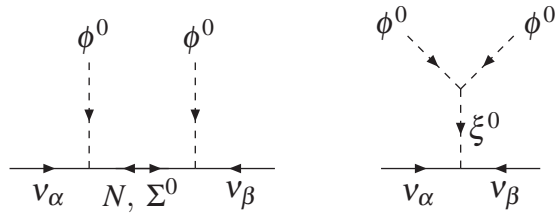
\begin{figure}[htb]
%\begin{center}
\begin{picture}(360,80)(0,0)
\ArrowLine(80,10)(110,10)
\ArrowLine(180,10)(150,10)
\ArrowLine(140,10)(110,10)
\ArrowLine(120,10)(150,10)
\DashArrowLine(110,50)(110,10)3
\DashArrowLine(150,50)(150,10)3
\ArrowLine(220,10)(250,10)
\ArrowLine(280,10)(250,10)
\DashArrowLine(250,40)(250,10)3
\DashArrowLine(230,60)(250,40)3
\DashArrowLine(270,60)(250,40)3
\Text(95,0)[]{$\nu_\alpha$}
\Text(130,0)[]{$N,~\Sigma^0$}
\Text(168,0)[]{$\nu_\beta$}
\Text(112,60)[]{$\phi^0$}
\Text(152,60)[]{$\phi^0$}
\Text(225,65)[]{$\phi^0$}
\Text(281,65)[]{$\phi^0$}
\Text(260,25)[]{$\xi^0$}
\Text(235,0)[]{$\nu_\alpha$}
\Text(268,0)[]{$\nu_\beta$}
\end{picture}
%\end{center}
\caption{Three tree-level realizations of seesaw Majorana neutrino mass.}
\end{figure}
The most common thinking regarding the seesaw origin of neutrino mass is 
to assume a heavy Majorana fermion singlet $N$ (version 1), the next most 
common is to use a heavy scalar triplet $(\xi^{++},\xi^+,\xi^0)$ (version 2), 
whereas the third option, i.e. that of a heavy Majorana fermion triplet 
$(\Sigma^+,\Sigma^0,\Sigma^-)$ \cite{flhj89} (version 3), has not received 
as much attention.  However, it may be relevant to a host of other issues in 
physics beyond the SM and is now being studied extensively.  I will review 
in this talk a number of such topics, including gauge-coupling unification 
in the SM, new U(1) gauge symmetry, and dark matter.

\section{Gauge-Coupling Unification}

It is well-known that gauge-coupling unification occurs for the minimal 
supersymmetric standard model (MSSM) but not the SM.  The difference can 
be traced to the addition of gauginos and higgsinos, transforming under 
$SU(3)_C \times SU(2)_L \times U(1)_Y$ as (8,1,0), (1,3,0), $(1,2,\pm 1/2)$,
and a second Higgs scalar doublet.  In particular, the contribution of 
the $SU(2)_L$ gaugino triplet is crucial in allowing the $SU(2)_L$ and 
$U(1)_Y$ gauge couplings to meet at high enough an energy scale to 
be acceptable for suppressing proton decay.  Since $\Sigma$ is exactly 
such a fermion triplet, it is not surprising that gauge-coupling 
unification in the SM may be achieved using it \cite{m05,bs07,df07,ms08} 
together with some other fields.

To understand how this works, consider the one-loop renormalization-group 
equations governing the evolution of the three gauge couplings with mass 
scale:
\begin{equation}
{1 \over \alpha_i (M_1)} - {1 \over \alpha_i (M_2)} = {b_i \over 2 \pi} 
\ln {M_2 \over M_1},
\end{equation}
where $\alpha_i = g_i^2/4 \pi$, and the numbers $b_i$ are determined by the 
particle content of the model between $M_1$ and $M_2$.  Since
\begin{equation}
\alpha_C(M_U) = \alpha_L(M_U) = (5/3)\alpha_Y(M_U) = \alpha_U 
\end{equation}
is required for unification, but not the actual numerical value of $\alpha_U$, 
only $b_Y-b_L$ and $b_L-b_C$ are important for this purpose.  These numbers 
are listed below for the SM, MSSM, and some other models.
\begin{table}[htb]
\caption{Gauge-coupling unification in the MSSM and other models.}
%\begin{center}
\begin{tabular}{|c|c|c|c|c|}
\hline 
Model & $b_Y-b_L$ & $b_L-b_C$ & new fermions & new scalars\\ 
\hline
SM & 7.27 & 3.83 & none & none \\ 
\hline
MSSM & 5.60 & 4.00 & (1,3,0), ~(8,1,0), ~(1,2,$\pm 1/2$) & (1,2,1/2) \\
\hline
Ref.~\cite{m05} & 5.27 & 3.83 & (1,3,0) & (1,3,0) $\times$ 2, ~(8,1,0) 
$\times$ 4 \\
\hline
Ref.~\cite{bs07,df07} & 5.60 & 3.00 & (1,3,0), ~(8,1,0) & (1,3,0), ~(8,1,0) \\
\hline
Ref.~\cite{ms08} & 5.87 & 4.33 & (1,3,0) & (1,2,1/2), ~(8,1,0) $\times$ 2 \\
\hline
\end{tabular}
%\end{center}
\end{table}
Focus only on those new particles which transform nontrivially under $SU(2)_L 
\times U(1)_Y$.  Let them be at the electroweak scale, then
\begin{equation}
\ln {M_U \over M_Z} \simeq {\sqrt{2} \pi^2 \over (b_Y-b_L) G_F M_W^2} 
\left( {3 \over 5 \tan^2 \theta_W} - 1 \right).
\end{equation}
Hence $M_U$ greater than about $10^{16}$ GeV implies $b_Y-b_L$ less than 
about 5.7.  In Refs.~\cite{bs07,df07}, an intermediate scale of about 
$10^8$ GeV is needed for the color octets.

\section{Phenomenology of $(\Sigma^+,\Sigma^0,\Sigma^-)$}

If $\Sigma$ exists at or below the TeV scale, then it has a rich phenomenology 
\cite{flhj89,bns07,fhs08,aa08} and may be probed at the Large Hadron Collider 
(LHC).  Unless there is a Higgs scalar triplet $(s^+,s^0,s^-)$ \cite{m05}, the 
mass splitting between $\Sigma^0$ and $\Sigma^\pm$ is radiative and comes 
from electroweak gauge interactions. It is positive and for large $m_\Sigma$, 
it approaches \cite{cfs06} $G_F M_W^3 (1 - \cos \theta_W)/\sqrt{2} \pi 
\simeq 168$ MeV, thus allowing the decay of $\Sigma^\pm$ to $\Sigma^0 \pi^\pm$ 
and $\Sigma^0 l^\pm \nu$.  Since $\Sigma$ also has Yukawa couplings to 
$(\nu_\alpha,l_\alpha)$ and $(\phi^+,\phi^0)$, the decays $\Sigma^\pm \to 
l^\pm h, ~\Sigma^0 \to \nu h$ are possible, as well as $\Sigma^\pm \to l^\pm Z, 
~\nu W^\pm$ and $\Sigma^0 \to \nu Z, ~l^\pm W^\mp$ through the mixing of 
$\Sigma^0$ with $\nu$, and $\Sigma^\pm$ with $l^\pm$,  unless they are 
forbidden by a symmetry, in which case $\Sigma^0$ is a dark-matter (DM) 
candidate \cite{m05,cfs06,ff08}. 

The production of $\Sigma$ is by pairs from quark fusion through the 
electroweak gauge bosons with a cross section of the order 1 fb for 
$m_\Sigma$ of about 1 TeV, and rising to more than $10^2$ fb if $m_\Sigma$ is  
300 GeV.  Each decay mode of $\Sigma$ has a huge SM background to contend 
with.  The best chance of digging out the signal is to look for 
charged-lepton final states. Copying Ref.~\cite{aa08}, the prognosis at 
the LHC for the 5$\sigma$ discovery of the particles responsible for the 
three versions of the seesaw mechanism is shown below. A dash means no 
such state. A cross means no such signal.
\begin{table}[htb]
\caption{Discovery potential at the LHC for seesaw 1,2,3.}
\begin{tabular}{|c|c|c|c|}
\hline 
final state & $m_N = 100$ GeV & $m_\xi = 300$ GeV & $m_\Sigma = 300$ GeV \\ 
\hline
6 leptons & -- & -- & $\times$ \\ 
\hline
5 leptons & -- & -- & 28 fb$^{-1}$ \\ 
\hline
$l^\pm l^\pm l^\pm l^\mp$ & -- & -- & 15 fb$^{-1}$ \\ 
\hline
$l^+l^+l^-l^-$ & -- & 19 fb$^{-1}$ & 7 fb$^{-1}$ \\ 
\hline
$l^\pm l^\pm l^\pm$ & -- & -- & 30 fb$^{-1}$ \\ 
\hline
$l^\pm l^\pm l^\mp$ & <180 fb$^{-1}$ & 3.6 fb$^{-1}$ & 2.5 fb$^{-1}$ \\ 
\hline
$l^\pm l^\pm$ & <180 fb$^{-1}$ & 17.4 fb$^{-1}$ & 1.7 fb$^{-1}$ \\ 
\hline
$l^+ l^-$ & $\times$  & 15 fb$^{-1}$ & 80 fb$^{-1}$ \\ 
\hline
$l^\pm$ & $\times$ & $\times$ & $\times$ \\
\hline
\end{tabular}
\end{table}

\section{Leptogenesis involving $(\Sigma^+,\Sigma^0,\Sigma^-)$}

Jsut as there are three seesaw mechanisms, the decays of the corresponding 
heavy particles $N$ \cite{fy86}, $(\xi^{++},\xi^+,\xi^0)$ \cite{ms98}, and 
$(\Sigma^+,\Sigma^0,\Sigma^-)$ \cite{ff08} are natural for generating a 
lepton asymmetry of the Universe, which gets converted \cite{krs85} 
into the present observed baryon asymmetry through sphalerons.  Just as 
$N$ may decay into leptons and antileptons because it is a Majorana 
fermion, the same is true for $\Sigma$.  Assuming three such triplets, 
successful leptogenesis requires \cite{ff08} the lightest to be heavier 
than about $10^{10}$ GeV, similar to that for the lightest $N$.  However, 
since $\Sigma$ has electroweak gauge interactions, the initial conditions 
for the Boltzmann equations are determined here through thermal equilibrium, 
which may not be as simple for $N$.

There is another interesting correlation.  The addition of three $(1,3,0)$ 
fermion triplets to the SM instead of just one will not lead to gauge-coupling 
unification unless all three are also roughly at the $10^{10}$ GeV scale 
\cite{ff08}.  Whereas other fields are still needed, such as those 
transforming under $(8,1,0)$, this is another argument for preferring 
$\Sigma$ over $N$. 

\section{New U(1) Gauge Symmetry}

Consider an extension of the SM to include a fermion triplet $(\Sigma^+,
\Sigma^0,\Sigma^-)$ per family as well as a new $U(1)_X$ gauge symmetry 
as listed below.
\begin{table}[htb]
\caption{Fermion content of proposed model.}
\begin{tabular}{|c|c|c|}
\hline 
Fermion & $SU(3)_C \times SU(2)_L \times U(1)_Y$ & $U(1)_X$ \\ 
\hline
$(u,d)_L$ & $(3,2,1/6)$ & $n_1$ \\ 
$u_R$ & $(3,1,2/3)$ & $n_2=(7n_1-3n_4)/4$ \\
$d_R$ & $(3,1,-1/3)$ & $n_3=(n_1+3n_4)/4$ \\
$(\nu,e)_L$ & $(1,2,-1/2)$ & $n_4 \neq -3n_1$ \\
$e_R$ & $(1,1,-1)$ & $n_5=(-9n_1+5n_4)/4$ \\
\hline
$(\Sigma^+,\Sigma^0,\Sigma^-)_R$ & $(1,3,0)$ & $n_6=(3n_1+n_4)/4$ \\
\hline
\end{tabular}
\end{table}
Remarkably \cite{m02,mr02,bd05}, $U(1)_X$ is free of all anomalies. 
For example, one can easily check that
\begin{equation}
6n_1^3 - 3n_2^3 - 3n_3^3 + 2n_4^3 - n_5^3 = 3 (3n_1+n_4)^3/64 = 3 n_6^3.
\end{equation}
Furthermore, it has been shown \cite{mr02} that if a fermion multiplet 
$(1,2p+1,0;n_6)$ per family is added to the SM, the only anomaly-free 
solutions for $U(1)_X$ are $p=0$ ($N$) for which the well-known $U(1)_{B-L}$ 
is obtained, and $p=1$ ($\Sigma$) as given above.

The new gauge boson $X$ may be accessible at the LHC.  In that case, its 
decay into quarks and leptons will determine the parameter $r=n_4/n_1$. In 
particular, the ratios
\begin{eqnarray}
{\Gamma (X \to t \bar{t}) \over \Gamma (X \to \mu \bar{\mu})} = 
{3(65 - 42r + 9r^2) \over 81 - 90r + 41r^2}, ~~~
{\Gamma (X \to b \bar{b}) \over \Gamma (X \to \mu \bar{\mu})} = 
{3(17 + 6r + 9r^2) \over 81 - 90r + 41r^2},
\end{eqnarray}
are especially good discriminators \cite{gm08}, 
as shown in Fig.~2 \cite{aem08}.

\input epsf
\begin{figure}[htb]
\centerline{
\epsfxsize=8cm
\leavevmode
\epsfbox{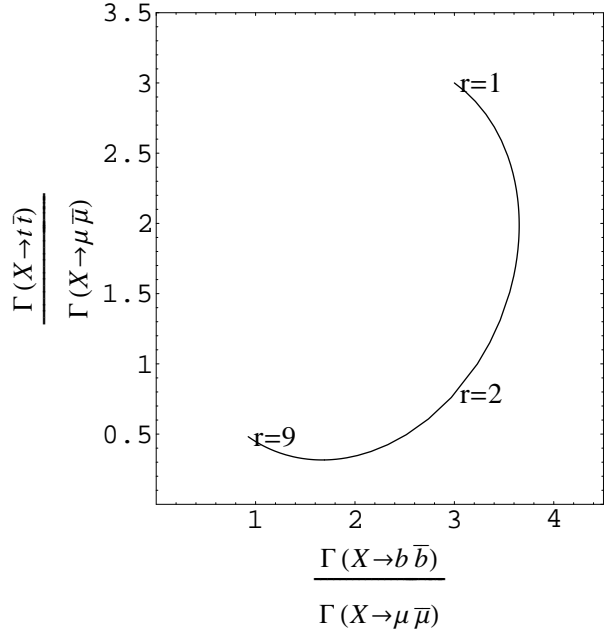}
%\end{center}
\vspace*{-5mm}
\caption{Plot of $\Gamma(X \to t \bar{t})/\Gamma(X \to \mu \bar{\mu})$ 
versus $\Gamma(X \to b \bar{b})/\Gamma(X \to \mu \bar{\mu})$.}
}
\end{figure}

The scalar sector of this $U(1)_X$ model consists of two Higgs doublets 
$\Phi_1 = (\phi_1^+,\phi_1^0)$ with charge $(9n_1-n_4)/4$ which couples 
to charged leptons, and $\Phi_2 = (\phi_2^+,\phi_2^0)$ with charge 
$(3n_1-3n_4)/4$ which couples to $up$ and $down$ quarks as well as to 
$\Sigma$.  To break the $U(1)_X$ gauge symmetry spontaneously, a singlet 
$\chi$ with charge $-2n_6$ is added, which also allows the $\Sigma$'s 
to acquire Majorana masses at the $U(1)_X$ breaking scale.  This specific 
two-Higgs doublet model is different from conventional studies where 
one doublet couples to $up$ quarks and the other to $down$ quarks and 
charged leptons.  The resulting detailed differences are verifiable at 
the LHC.  

In general, there is $Z-X$ mixing in their mass matrix, but it 
must be very small to satisfy present precision electroweak measurements.  
The condition for zero $Z-X$ mass mixing is 
$v_1^2/v_2^2 = 3(n_4-n_1)/(9n_1-n_4)$, which 
requires $1 < n_4/n_1 < 9$. 
Low-energy precision measurements of SM physics also constrain the 
contributions of this $U(1)_X$.  Let $n_1^2 + n_4^2$ be normalized to one, 
and $\tan \phi = n_4/n_1$, then the 95\% confidence-level lower bound on 
$M_X/g_X$ is shown in Fig.~3 \cite{aem08}, assuming zero $Z-X$ mixing so 
that there is no constraint coming from measurements at the $Z$ resonance.
Thus only the range $1 < r < 9$, i.e. $\pi/4 < \phi < 1.46$ is actually 
allowed.
\vspace*{1cm}

\begin{figure}[htb]
\centerline{
\epsfxsize=10cm
\leavevmode
\epsfbox{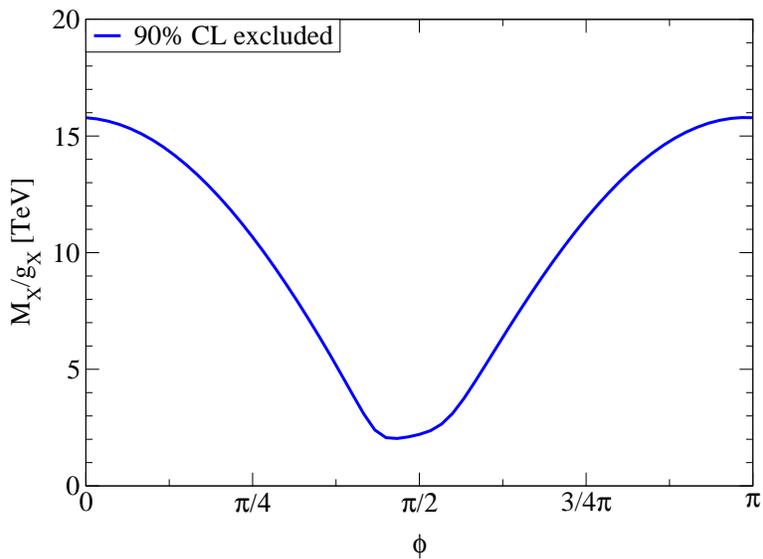}
%\end{center}
\vspace*{-5mm}
\caption{Lower bound on $M_X/g_X$ versus $\phi$.}
}
\end{figure}

\newpage
\section{Scotogenic radiative neutrino mass}

There are also three generic one-loop radiative mechanisms \cite{m98} for 
neutrino mass.  An intriguing possibility is that the 
particles in the loop are distinguished from those of the SM by a 
$Z_2$ discrete symmetry.  The simplest realization \cite{m06} is to 
add a second scalar doublet $(\eta^+,\eta^0)$ \cite{dm78} as well as 
three fermion singlets $N$, and let them be odd under $Z_2$ with all SM 
particles even.  Clearly, $\Sigma$ may be chosen \cite{ms08} instead of 
$N$ and a radiative seesaw neutrino mass is generated as shown in Fig.~4.
\begin{figure}[htb]
%\begin{center}
\begin{picture}(360,100)(0,0)
\ArrowLine(90,10)(130,10)
\ArrowLine(180,10)(130,10)
\ArrowLine(180,10)(230,10)
\ArrowLine(270,10)(230,10)
\DashArrowLine(155,85)(180,60)3
\DashArrowLine(205,85)(180,60)3
\DashArrowArc(180,10)(50,90,180)3
\DashArrowArcn(180,10)(50,90,0)3
\Text(110,0)[]{$\nu_\alpha$}
\Text(250,0)[]{$\nu_\beta$}
\Text(180,0)[]{$\Sigma^0_i$}
\Text(135,50)[]{$\eta^0$}
\Text(230,50)[]{$\eta^0$}
\Text(150,90)[]{$\phi^{0}$}
\Text(217,90)[]{$\phi^{0}$}
\end{picture}
%\end{center}
\caption{One-loop generation of seesaw neutrino mass.}
\end{figure}
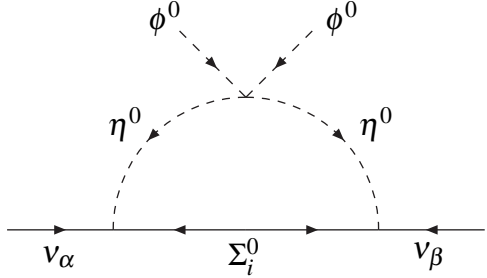
The allowed quartic scalar term $(\lambda_5/2)(\Phi^\dagger \eta)^2 + H.c.$ 
is necessary for this mechanism to work.  It also splits the complex scalar 
field $\eta^0$ into two mass eigenstates: Re($\eta^0$) and Im($\eta^0$), 
resulting in
\begin{equation}
({\cal M}_\nu)_{\alpha \beta} = \sum_i {h_{\alpha i} h_{\beta i} M_i \over 16 \pi^2} 
\left[ {m_R^2 \over m_R^2-M_i^2} \ln {m_R^2 \over M_i^2} - {m_I^2 \over 
m_I^2-M_i^2} \ln {m_I^2 \over M_i^2} \right],
\end{equation}
where $m_R^2-m_I^2 = 2 \lambda_5 v^2$ and $M_i$ are the $\Sigma$ masses.  The 
lighter one of Re($\eta^0$) and Im($\eta^0$) is then a good candidate 
\cite{bhr06,lnot07,glbe07,cmr07} for dark matter (DM).  Neutrino mass may 
then be called scotogenic, i.e. being caused by darkness \cite{m08}.

\section{$\Sigma^0$ as dark matter}

In Ref.~\cite{m06}, the lightest $N$ may also be a DM candidate 
\cite{kms06,akrsz08}, but then its only interaction is with $(\nu_\alpha 
\eta^0 - l_\alpha \eta^+)$ and these couplings have to be rather large 
to obtain the requisite DM relic abundance.  In that case, 
flavor-changing radiative decays such as $\mu \to e \gamma$ are generically 
too big and require delicate fine tuning among the masses and couplings of $N$ 
to be consistent with data.

If $\Sigma^0$ is selected as dark matter, then it can annihilate with itself 
and coannihilate with the slightly heavier $\Sigma^\pm$ through electroweak 
gauge interactions to account for the correct relic abundance.  Its Yukawa 
couplings may then be appropriately small, not to upset the constraints 
from $\mu \to e \gamma$, etc. Using the method developed in Ref.~\cite{gs91} 
to take coannihilation into account, and the various cross sections times 
the absolute value of the relative velocity of the DM particles, namely
\begin{eqnarray}
&& \sigma(\Sigma^0 \Sigma^0) |v| \simeq {2 \pi \alpha_L^2 \over m_\Sigma^2}, 
~~~ \sigma(\Sigma^\pm \Sigma^\pm) |v| \simeq {\pi \alpha_L^2 \over m_\Sigma^2}, 
\\ && \sigma(\Sigma^+ \Sigma^-) |v| \simeq {37 \pi \alpha_L^2 \over m_\Sigma^2}, 
~~~ \sigma(\Sigma^0 \Sigma^\pm) |v| \simeq {29 \pi \alpha_L^2 \over m_\Sigma^2}, 
\end{eqnarray}
$m_\Sigma$ is estimated \cite {ms08} to be in the range 2.28 to 2.42 TeV 
to reproduce the observed data $\Omega h^2 = 0.11 \pm 0.006$ \cite{dm03} 
for its relic abundance.  Note that the presence of $\Sigma^\pm$ is 
important for having a large enough effective annihilation cross section 
for this to work and that the only free parameter here is $m_\Sigma$. The 
validity of $\Sigma^0$ as dark matter depends only on $Z_2$ and not on 
whether it is the source of radiative neutrino mass.

\section{$\Sigma$ as lepton and $N$ as baryon}

Assuming neutrino masses come from $\Sigma$, an intriguing possibility exists 
that the heavy fermion singlet $N$ may in fact be a baryon \cite{m88,mr02-1,
m08-2,m08-3,m08-4}.  The crucial ingredient for this unconventional 
identification is the existence of a scalar diquark $\tilde{h} \sim (3,1,-1/3)$ 
with baryon number $B = -2/3$ so that the Yukawa couplings $u d \tilde{h}$, 
$u^c d^c \tilde{h}^*$, and $d^c N \tilde{h}$ are allowed, thereby making 
$N$ a baryon $(B=1)$.  Since $N$ is a gauge singlet, it is also allowed a 
large Majorana mass.  Hence additive $B$ breaks to multiplicative $(-)^{3B}$ 
and the decays of the lightest $N$ to $u d d$ and $\bar{u} \bar{d} \bar{d}$ 
through $\tilde{h}$ would produce a baryon asymmetry in the early Universe. 
Below the mass scale of $m_N$, baryon number is again additively conserved, 
allowing this pure $B$ asymmetry to be converted into a conserved $B-L$ 
asymmetry through the electroweak sphalerons, in analogy to the well-known 
scenario of leptogenesis \cite{dnn08}.

\section{Conclusion}

Using the fermion triplet $(\Sigma^+,\Sigma^0,\Sigma^0)$ as the seesaw anchor 
for neutrino masses (version 3), many new and interesting possibilities of 
physics beyond the SM exist.  It may be the missing link for 
gauge-coupling unification in the SM without going to the MSSM.  As a 
result, the phenomenological landscape at the TeV scale may change 
significantly and be verifiable at the LHC, where $\Sigma$ itself is much 
easier to detect than its singlet counterpart $N$.  There may also be an 
associated neutral gauge boson, corresponding to an anomaly-free $U(1)_X$, 
whose decays into quarks and leptons are predicted as a function of a 
single parameter $r = n_4/n_1$.  Furthermore, $\Sigma$ may be the source of 
scotogenic radiative neutrino masses and be a dark-matter candidate itself, 
with a mass around 2.35 TeV.  Other recent discussions of fermion triplets 
are found in Refs.~\cite{f07,abbgh08,m08-1,moy08,f08}.

\section{Acknowledgements} I thank Abdel Perez-Lorenzana and the other 
organizers of the XIII Mexican School of Particles and Fields for their 
great hospitality and a stimulating meeting in San Carlos.  This work was 
supported in part by the U.~S.~Department of Energy under Grant 
No.~DE-FG03-94ER40837.

%%%%%%%%%%%%%%%%%%%%%%%%%%%%%%%%%%%%%%%%%%%%%%%%
%% BACKMATTER
%%%%%%%%%%%%%%%%%%%%%%%%%%%%%%%%%%%%%%%%%%%%%%%%



\bibliographystyle{aipproc}   % if natbib is available


%%%%%%%%%%%%%%%%%%%%%%%%%%%%%%%%%%%%%%%%%%%
%% The following lines show an example how to produce a bibliography
%% without the help of the BibTeX program. This could be used instead
%% of the above.
%%%%%%%%%%%%%%%%%%%%%%%%%%%%%%%%%%%%%%%%%%%

\end{document}